\documentclass{article}
\usepackage{latexsym}
\usepackage{epsfig}
\begin{document}

\title{Gravitational Radiation Damping\\ and the Three-Body Problem}
\author{Zachary E. Wardell\thanks{e-mail: zew554@mizzou.edu}
\\Department of Physics and Astronomy\\
University of Missouri-Columbia\\Columbia, Missouri 65211, USA}

\maketitle
\begin{abstract}
A model of three-body motion is developed which includes the effects of
gravitational radiation reaction. 
The radiation reaction due to the emission of gravitational waves is the 
only post-Newtonian effect that is included here. 
For simplicity, all of the motion is taken to be planar.
Two of the masses are viewed as a binary system and the third mass, whose
motion will be a fixed orbit around the center-of-mass of the binary
system, is viewed as a perturbation.
This model aims to describe the motion of a relativistic binary pulsar
that is perturbed by a third mass. 
Numerical integration of this simplified model
reveals that
given the right initial conditions and parameters one can see resonances.
These $(m,n)$ resonances are defined by the resonance condition, $m\omega=
2n\Omega$, where $m$ and $n$ are relatively prime integers and $\omega$ and
$\Omega$ are the angular frequencies of the binary orbit and third mass orbit
(around the center-of-mass of the binary), respectively. The resonance condition
consequently fixes a value for the semimajor axis of the binary orbit for the
duration of the resonance; therefore, the binary energy remains constant on the
average while its angular momentum changes during the resonance.
\end{abstract}

\section{INTRODUCTION}

In 1975, Hulse and Taylor found through the timing observations of 
binary pulsar PSR B1913+16 that the semimajor axis decayed at nearly the
rate
predicted by general relativity for the emission of gravitational radiation.
This has served as significant indirect
evidence of gravitational radiation reaction force as the
result of energy balance due to the fact that gravitational waves leave the
system. It has also shown the importance of the
pulsar in probing strong gravity. Because of the universal nature of 
gravity, a correct analysis of astrophysical systems
must involve the gravitational interaction between all of the masses in 
the environment plus other non-Newtonian effects. One would expect that over time,
evidence of neighboring masses would appear in the timing observations of
binary pulsars. 

This study puts forth a model of three bodies, two of which constitute
a relativistic binary pulsar and the third is a perturbing mass.
This investigation extends the work of Chicone, Mashhoon, and
Retzloff (1996a,1996b,1997a,1997b), which originally dealt with a binary system 
perturbed by
normally incident gravitational waves. As in these previous models, the
model derived here will be searched for resonances. In the following
analysis, an $(m,n)$ resonance will occur when the relation $m\omega=2n\Omega$ is
satisfied, where $m$ and $n$ are relatively prime integers, $\omega$ is the
frequency of the binary system, and $\Omega$ is the frequency of the orbit
of the third mass around the center-of-mass of the binary. Such resonances
are physically noteworthy because they correspond to events where the
collapse of the semimajor axis of the binary halts. That is, on average it 
stays fixed.

The N-body problem in general
relativity theory is rather complicated; therefore, to capture the main effects
of three bodies plus gravitational radiation reaction one can start with a
classical three-body system with gravitational radiation damping as a linear
perturbation. That is, in this work all post-Newtonian (relativistic)
corrections will be neglected except for gravitational radiation reaction
that will be treated in the quadrupole approximation. 
For the sake of simplicity, the internal structure of the masses will be 
neglected so that one is in effect dealing with three Newtonian point masses.

Two astrophysical
systems that may demonstrate measurable behavior based on this
model are the relativistic binary pulsar near a massive third mass and a planet
in an orbit around a binary pulsar. While this study will focus on the former case
the latter proves to be a timely point of interest. Astronomers 
have in the last six years found more than 50 stars which include at least one 
planet. 
The present number of stars that are currently being observed for planets is
1000 (Cameron, 2001). Pulsars have also proven to be detectable sources
that show evidence of extrasolar planets. Finding a pulsar planet, where the 
pulsar is also part
of a binary system is a possibility. Microlensing technique in its extrasolar 
planet search has turned up a possible planet with two suns (Bennet et al., 1999).
Another possibility would be the future detection of a binary
pulsar in a globular cluster.  A binary pulsar system that is
situated in an approximately spherical cluster would be effectively the same as
a binary attracted to a third mass equal to the mass of the matter inside the 
binary's orbit.

The timing of radio pulses from pulsars has opened up a rich and 
exciting branch of astronomy and astrophysics.
Analysis of pulsar signals over time offers insight into the gravitational
environment such as the possible presence of a companion mass.
Hence, pulsar astronomy offers a fruitful method of detection for the
astrophysical processes discussed in this paper. On the frontiers of astronomy,
interferometric gravitational wave detectors based on Earth such as LIGO, 
VIRGO, GEO, AND TAMA aim to detect gravitational waves from known inspiraling
binary systems. The shortening orbital separation that would arise as the
result of energy loss via gravitational wave emission would result in an
increasing frequency in the orbit. The gravitational wave signal would have 
twice this frequency (Blanchet, 2002).
Thus, the predicted effect of a binary system's capture into resonance would 
leave its imprint in terms of the detected gravitational waves. Namely, the
interval when the distance between the members of the binary stays fixed
on average would give a gravitational wave signal that on the average would
be of near constant frequency.
Furthermore, the next generation of planned space-based detectors like LISA 
will broaden the scope of gravitational wave detection.

The initial approach to this problem in Section 2
involves setting up the equations of motion for the three masses with
the influence of gravitational radiation reaction included. The radiation
reaction force is understood to be very small, so it is viewed as a
perturbation. The model is then simplified in Section 3 to make it
more amenable to analysis. The results of the numerical integration of the
simplified model are presented in Section 4. Next, section 5 expands upon 
the numerical results with a discussion about resonance. 
Section 6 contains concluding remarks.

The nonlinear disposition of the equations of motion for the relative
motion of the binary system requires 
analysis that is suited for nonlinear behavior. Chicone, Mashhoon,
and Retzloff (1997b) developed such an averaging method, that elucidates 
details about the structure of the orbit of a nonlinear system 
, especially when a resonance occurs. The system developed 
and studied here is suited for
further analysis, in particular the application of the method mentioned above.

\section{EQUATIONS OF MOTION}

The general equations of motion for the three-body problem with
gravitational radiation damping are:

\begin{equation}
m_{1} \frac{d^{2}x^{i}_{1}}{dt^{2}}+\frac{Gm_{1}m_{2}(x^{i}_{1}-x^{i}_{2})}
{|\textbf{x}_{1}-\textbf{x}_{2}|^{3}} = -\frac{Gm_{1}m_{3}(x^{i}_{1}-
x^{i}_{3})}{|\textbf{x}_{1}-\textbf{x}_{3}|^{3}} - \frac{2G}{15c^{5}}
m_{1} \frac{d^{5}D_{ij}}{dt^{5}}x^{j}_{1},
\end{equation}

\begin{equation}
m_{2} \frac{d^{2}x^{i}_{2}}{dt^{2}}+\frac{Gm_{1}m_{2}(x^{i}_{2}-x^{i}_{1})}
{|\textbf{x}_{1}-\textbf{x}_{2}|^{3}} = -\frac{Gm_{2}m_{3}(x^{i}_{2}-
x^{i}_{3})}{|\textbf{x}_{2}-\textbf{x}_{3}|^{3}} - \frac{2G}{15c^{5}}
m_{2} \frac{d^{5}D_{ij}}{dt^{5}}x^{j}_{2},
\end{equation}

\begin{equation}
m_{3} \frac{d^{2}x^{i}_{3}}{dt^{2}}+\frac{Gm_{3}m_{1}(x^{i}_{3}-x^{i}_{1})}
{|\textbf{x}_{3}-\textbf{x}_{1}|^{3}} = -\frac{Gm_{2}m_{3}(x^{i}_{3}-
x^{i}_{2})}{|\textbf{x}_{3}-\textbf{x}_{2}|^{3}} - \frac{2G}{15c^{5}}
m_{3} \frac{d^{5}D_{ij}}{dt^{5}}x^{j}_{3},
\end{equation}
where the quadrupole moment tensor for the three-body system is:
\begin{equation}
D_{ij}=m_{1}(3x^{i}_{1}x^{j}_{1} - \delta_{ij}x_{1}^{2})+
m_{2}(3x^{i}_{2}x^{j}_{2} - \delta_{ij}x_{2}^{2}) +
m_{3}(3x^{i}_{3}x^{j}_{3} - \delta_{ij}x_{3}^{2}).
\end{equation}

Each equation includes the familiar Newtonian gravitational interactions
between the masses plus the Newtonian gravitational radiation reaction
force. This damping force results from the emission of gravitational waves
by the system of masses. There is a loss of energy in the system that 
results from energy carried away by the waves. According to general relativity,
the energy of gravitational waves leaving the system is given in the quadrupole
approximation by

\begin{equation}
\frac{dE}{dt} =\frac{G}{45c^{5}} \frac{d^{3}D_{ij}}
{dt^{3}} \frac{d^{3}D^{ij}}{dt^{3}}, 
\end{equation}
where $E$ is the amount of gravitational radiation energy emitted by
the system (Landau and Lifshitz, 1971).

Energy conservation requires
that this loss be reflected in the equations of motion; therefore, one can also 
arrive at this rate of energy loss by way of mechanics. If
one multiplies each equation of motion by its time derivative of position and 
adds the
three equations, one arrives at the following expression: 
\begin{equation}
\frac{d\mathcal{E}}{dt}=-\frac{G}{45c^{5}}\frac{d^{3}D_{ij}}{dt^{3}}
\frac{d^{3}D^{ij}}{dt^{3}},
\end{equation}
where $\mathcal{E} = \mathcal{E}_{N} + \mathcal{E}_{S}$. Here $\mathcal{E}_{N}$
is the total Newtonian energy of the three-body system. That is,
\begin{equation}
\mathcal{E}_{N}=\sum_{i=1}^{3} \frac{1}{2}m_{i}v_{i}^{2}-\frac{1}{2}\sum_{i\neq j}
\frac{Gm_{i}m_{j}}{|\textbf{x}_{i}-\textbf{x}_{j}|},
\end{equation}
and $\mathcal{E}_{S}$ is the gravitational ``Schott'' energy,
\begin{equation}
\mathcal{E}_{S}=\frac{G}{45c^{5}}
\left(\frac{d^{4}D_{ij}}
{dt^{4}}\frac{dD^{ij}}{dt} - \frac{d^{3}D_{ij}}{dt^{3}}
\frac{d^{2}D^{ij}}{dt^{2}}\right),
\end{equation}
that is analogous to the Schott term in electrodynamics (Schott, 1912).

For quasi-periodic motions, the change in the Schott energy over a period turns
out to be of higher order and can be neglected.
Hence there is agreement between the two different methods as to what the loss
of energy should be since the right-hand side of (5) is equal and opposite to
the right-hand side of (6). Comparison of the general relativistic and classical
methods of calculating rate of change of angular momentum yields agreement in a
similar way. Thus in the quadrupole approximation for the emission of gravitational
waves, the three-body system loses energy and angular momentum to the radiation
field. Moreover, adding equations (1)-(3) and making the
following definition
\begin{equation}
\textbf{Z}=\frac{m_{1}\textbf{x}_{1} + m_{2}\textbf{x}_{2} + m_{3}\textbf{x}_{3}}
{m_{1}+m_{2}+m_{3}}
\end{equation}
for the center-of-mass of the system we find
\begin{equation}
\frac{d^{2}Z^{i}}{dt^{2}} + \frac{2G}{15c^{5}}\frac{d^{5}D_{ij}}{dt^{5}}Z^{j}=0.
\end{equation}
Let us note that $\textbf{Z}=0$ is a solution of this equation, so that if the
center-of-mass of the whole system is initially at rest at the origin of
coordinates, it will remain so in the quadrupole approximation under consideration
here. This is consistent with the fact that gravitational waves do not carry
away linear momentum in the quadrupole approximation.

Solutions to the proposed three-body problem, equations (1)-(3),
by way of numerical integration
offer some insight into the behavior of the system that has no known
solution in closed form. To integrate such a higher-order system
numerically, it is not sufficient to specify the positions and
velocities of the three masses at some initial instant of time. In fact,
it is known that such systems suffer from the existence of runaway modes that
inevitably will lead to divergent results. It is therefore necessary to replace
the system (1)-(3) by second-order equations of motion via iterative
reduction as explained in a recent paper by Chicone, Kopeikin, Mashhoon, and
Retzloff (2001). This procedure involves the repeated substitution of the
equations of motion (1)-(3) in the evaluation of $d^{5}D_{ij}/dt^{5}$ and
the subsequent reduction of the resulting system to one of second-order
equations for the motion of the system (by dropping higher-order terms). 
The resulting system would be
appropriate for numerical integration; however, it would have a rather
complicated form. Therefore, we resort to certain simplifications in this
first treatment of the classical three-body problem that takes gravitational
radiation reaction into account.

The main physical model which will be under investigation in this study
involves two masses, which represent the binary, whose center-of-mass
orbits a massive third body. Since the mass of the third body is assumed to
be much
larger than the mass of the binary, one can take $\textbf{x}_{3}$ to be the 
origin of the coordinate system. This is because the center-of-mass of the
entire system is very close to the location of the third mass. To simplify
matters, we set $\mathbf{x}_{3} = 0$; moreover, the distance
of the binary from the third mass is taken to be much larger than the 
semimajor axis of the binary system.

This model could approximate a relativistic
binary pulsar system that orbits the center-of-mass of a globular cluster.
The outer shell of stars would remain 'unseen' by the binary since a globular 
cluster is nearly spherical, and the interior stars would serve as a third mass 
located at the center-of-mass of the cluster.

\section{SIMPLIFIED MODEL}

The following equations of motion result from the approximations made in
the model:
\begin{equation}
m_{1} \frac{d^{2}x_{1}^{i}}{dt^{2}} = -\frac{G m_{3}
m_{1}x_{1}^{i}}{|\textbf{x}_{1}|^{3}} +\frac{G m_{1}
m_{2}(x_{2}^{i}-x_{1}^{i})}{|\textbf{x}_{2} -
\textbf{x}_{1}|^{3}} - \frac{2 m_1 G}{15c^{5}} \frac{d^{5}
D_{ij}}{dt^{5}} x_{1}^{j},
\end{equation}
\begin{equation}
m_{2} \frac{d^{2}x_{2}^{i}}{dt^{2}} = -\frac{G m_{3}
m_{2}x_{2}^{i}}{|\textbf{x}_{2}|^{3}} - \frac{G
m_{1}m_{2}(x_{2}^{i}-x_{1}^{i})}{|\textbf{x}_{2}-\textbf{x}_{1}|^{3}} -
\frac{2 m_{2} G}{15 c^{5}} \frac{d^{5} D_{ij}}{dt^{5}} x_{2}^{j}.
\end{equation}
Here $m_{1}$ and $m_{2}$ are masses of the members of the binary,
$m_{3}$ is the mass of the perturbing third body, and
$D_{ij} = m_{1} (3x_{1}^{i}x_{1}^{j}-\delta_{ij}x_{1}^{2}) +
m_{2}(3x_{2}^{i}x_{2}^{j}- \delta_{ij}x_{2}^{2})$ is the quadrupole moment
tensor. 

In order to elucidate the relevant dynamics of the system, it is helpful
to change the coordinates of the system. The relative and center-of-mass 
coordinates of the binary can be defined as:

\begin{equation}
r^{i}=x_{2}^{i}-x_{1}^{i}, \ \  X^{i} = \frac{m_{1}x_{1}^{i}+m_{2}x_{2}^{i}}{M},
\end{equation}
where $M=m_{1}+m_{2}$.
The equation of relative motion is then
\begin{equation}
\frac{d^{2}r^{i}}{dt^{2}} = - \frac{G Mr^{i}}{|\textbf{r}|^{3}}
-
Gm_{3}\left(\frac{x_{2}^{i}}{|\textbf{x}_{2}|^{3}}-\frac{x_{1}^{i}}
{|\textbf{x}_{1}|^{3}}\right)- \frac{2 G}{15 c^{5}} \frac{d^{5} D_{ij}}
{dt^{5}}r^{j},
\end{equation}
while the equation of motion of the center-of-mass of the binary is
\begin{equation}
\frac{d^{2} X^{i}}{dt^{2}} = -\frac{Gm_{3}}{M}
\left(\frac{m_{1}x_{1}^{i}}{|\textbf{x}_{1}|^{3}} +
\frac{m_{2}x_{2}^{i}}{|\textbf{x}_{2}|^{3}}\right) - \frac{2 G}{15 c^{5}}
\frac{d^{5}
D_{ij}}{dt^{5}} X^{j}.
\end{equation}
One can show that the quadrupole moment tensor can be written in terms of the
relative and center-of-mass variables as
$D_{ij}=\mu(3r^{i}r^{j}-\delta_{ij}r^{2})+M(3X^{i}X^{j}-
\delta_{ij}X^{2})$, where
$\mu=m_{1}m_{2}/M$ is the reduced mass of the binary system,
$r=|\mathbf{r}|$, and $X=|\mathbf{X}|$.
Replacing $\mathbf{x}_{1}$ and $\mathbf{x}_{2}$ in (14)-(15) with
$x_{1}^{i}=X^{i}-\gamma_{2} r^{i}$ and $x_{2}^{i}=X^{i}+\gamma_{1}r^{i}$
and expanding terms in powers of $r/X$, one obtains
\begin{equation}
\frac{x_{1}^{i}}{|\textbf{x}_{1}|^{3}}=\frac{X^{i}-\gamma_{2}
r^{i}}{|\textbf{X}-\gamma_{2}\textbf{r}|^{3}}=
\frac{X^{i}}{X^{3}} -
\frac{\gamma_{2}}{X^{3}}\left(r^{i}-\frac{3(\textbf{X}
\cdot \textbf{r}) X^{i}}{X^{2}}\right)+... ,
\end{equation}

\begin{equation}
\frac{x_{2}^{i}}{|\textbf{x}_{2}|^{3}}=\frac{X^{i}+\gamma_{1} r^{i}}
{|\textbf{X}+\gamma_{1} \textbf{r}|^{3}}=\frac{X^{i}}{X^{3}}
+\frac{\gamma_{1}}{X^{3}}
\left(r^{i}-\frac{3(\textbf{X}\cdot \textbf{r})X^{i}}
{X^{2}}\right)+...  ,
\end{equation}
where $\gamma_{1} = m_{1}/M$ and $\gamma_{2}=m_{2}/M.$

The following equations result when terms to linear order in $r/X$ are 
kept in the above expansions:

\begin{equation}
\frac{d^{2}r^{i}}{dt^{2}} = - \frac{GM}{r^{3}}r^{i} 
-\frac{G m_{3}}{X^{3}} K_{ij} r^{j}
- \frac{2 G}{15 c^{5}} \frac{d^{5} D_{ij}}{dt^{5}} r^{j},
\end{equation}

\begin{equation}
\frac{d^{2}X^{i}}{dt^{2}} = -\frac{Gm_{3}}{X^{3}} X^{i}
- \frac{2G}{15c^{5}} \frac{d^{5}D_{ij}}{dt^{5}} X^{j},
\end{equation}
where
\begin{equation}
K_{ij}=\left(\delta_{ij} - \frac{3X^{i}X^{j}}{X^{2}}\right)
\end{equation}
is the reduced tidal matrix. 

Thus the problem reduces to the coupled system (18)-(19) for the motion of
the binary system. To simplify the equations further, it will be assumed in
what follows that the center-of-mass motion is circular in the absence of
radiation reaction. Substituting such a solution for (19) in the equation of
relative motion (18), one finds that to lowest order in the perturbations the
center-of-mass motion can be taken to be simply circular in (18). This 
circular motion contributes to the radiation reaction term in (18); however,
this contribution can also be neglected if

\begin{equation}
\left(\frac{X}{r}\right)^{11/2}\left(\frac{m_{1}}{m_{3}}\right)
\left(\frac{m_{2}}{m_{3}}\right)
\left(\frac{m_{1}+m_{2}}{m_{3}}\right)^{1/2} >> 1.
\end{equation}
Assuming this inequality, the equation of relative motion reduces to
\begin{equation}
\frac{d^{2}r^{i}}{dt^{2}} = - \frac{GM}{r^{3}}r^{i} 
-\frac{G m_{3}}{X^{3}} K_{ij} r^{j}
- \frac{2 G \mu}{15 c^{5}} \frac{d^{5} \tilde{D}_{ij}}{dt^{5}} r^{j},
\end{equation}
where $\tilde{D}_{ij} = 3r^{i}r^{j}-\delta_{ij}r^{2}.$ The rest of this paper is
devoted to the study of this equation that describes the relative motion of two 
bodies under mutual gravitational
attraction, tidal interaction with a large third mass, and gravitational
radiation damping. These are the combined effects of interest here. This model can, in effect,
represent a second scenario as well, i.e. a binary system orbited by a
distant small mass $m_{3}$ (see Appendix A).

It is convenient to transform equation (22) into dimensionless form. To
this end, let all lengths and temporal variables be measured in units of
$R_{0}$ and $T_{0}$, respectively. Moreover, one assumes that these units
are related to the unperturbed motion of the binary such that
$G M T_{0}^{2}=R_{0}^{3}$. Thus, letting $r^{i} \rightarrow
R_{0}\hat{r}^{i}$, $X^{i} \rightarrow R_{0} \hat{X}^{i}$, and
$t \rightarrow T_{0} \hat{t}$ in equation (22), it reduces to the form

\begin{equation}
\frac{d^{2}\hat{r}^{i}}{d\hat{t}^{2}} = -
\frac{\hat{r}^{i}}{\hat{r}^{3}}
-\Omega^{2}  K_{ij}\hat{r}^{j}
- \delta \frac{d^{5} \hat{\tilde{D}}_{ij}}{d\hat{t}^{5}}\hat{r}^{j},
\end{equation}
where the hats will be dropped in what follows for the sake of simplicity.
Here
\begin{equation}
\Omega^{2} =
\frac{Gm_{3}T_{0}^{2}}{X^{3}}=\frac{m_{3}}{M}\left(\frac{R_{0}}
{X}\right)^{3},
\end{equation}

\begin{equation}
\delta= \frac{2G\mu}{15c^{5}} \frac{R_{0}^{2}}{T_{0}^{3}},
\end{equation}
and the center-of-mass motion is taken to be a circle in the $(x,y)$-plane
such that $(K_{ij})$ has the form
\begin{equation}
(K_{ij}) =
\left[
\begin{array}{lll}
-\frac{1}{2}-\frac{3}{2}\cos{2 \Omega t} & -\frac{3}{2} \sin{2 \Omega t}
& 0 \\
-\frac{3}{2} \sin{2 \Omega t} & -\frac{1}{2}+\frac{3}{2}\cos{2 \Omega t}
&0 \\
0 & 0 & 1
\end{array}
\right].
\end{equation}

If the length unit $R_{0}$ were chosen to be the semimajor axis of the binary,
the corresponding initial period would be $2 \pi T_{0}$.
A relativistic binary pulsar such as PSR B1913+16 has an orbital period of
nearly eight hours (Lyne and Graham-Smith, 1998). The data regarding this
Hulse-Taylor binary pulsar when applied
to the dimensionless formula (25) give $ \delta \simeq 10^{-16}$. 
This value of $\delta$ is also approximately valid for the relativistic binary
pulsar PSR B1534+12 (Stairs et al., 1998). Moreover, the tidal perturbation 
term is also assumed to be very small, i.e. $\Omega^{2}<<1$. It proves useful
to replace $\Omega^{2}$ by a free parameter $\epsilon << 1$ in what
follows. Therefore, equation (22) will be replaced by

\begin{equation}
\frac{d^{2}r^{i}}{dt^{2}} = -
\frac{r^{i}}{r^{3}}
-\epsilon  K_{ij}r^{j}
- \delta \frac{d^{5} \tilde{D}_{ij}}{dt^{5}}r^{j}.
\end{equation}
The mathematical results derived from (27) will then apply to the physical
situation at hand once $\epsilon = \Omega^{2}$.

Finally, the order of equation (27) needs to be reduced in order to avoid a
singular perturbation problem. To this end, in taking derivatives of
$\tilde{D}_{ij}$ 
substitutions for $\ddot{r}^{i}$ are made with the equation
of motion (27), and terms to desired order are kept. 
In this case, we reduce (27) to a second-order equation that is linear in
$\epsilon$ and $\delta$.
See Appendix B for a complete derivation. What results is

\begin{equation}
\frac{d^{2}r^{i}}{dt^{2}} = -
\frac{r^{i}}{r^{3}}
-\epsilon  K_{ij}r^{j}-\delta R^{i}.
\end{equation}
Here $R^{i}$ is the reduced radiation reaction term in Cartesian
coordinates and is given by
\begin{equation}
R^{i}= \left(- \frac{24}{r} - 180\dot{r}^{2}
+72v^{2} \right)\frac{v^{i}}
{r^{3}} +
\left(-\frac{8}{r}+300\dot{r}^{2}-216v^{2}
\right) \frac{rr^{i}}{r^{4}},
\end{equation}
where as before $r=|\mathbf{r}|$, $v^{i}=\dot{r}^{i}$ and $v=|\mathbf{v}|$.

\section{NUMERICAL RESULTS}

What results is a nonlinear two-dimensional second-order ODE that
describes the relative motion between the two members of the binary.
Because of the nonlinearity, a numerical solution is sought at this
stage to elucidate the behavior of the system. The numerical integrator
used in this study was furnished by MATHEMATICA. To produce a trajectory
of the system, one needs values for the parameters $\epsilon, \delta,$
and $\Omega$ plus the necessary initial conditions. For the purposes of
generating numerical solutions for inspection, it becomes advantageous
to generalize the system as a dynamical system at the expense of adherence
to a strictly physical model. This relaxation of the parameter space 
allows one to more freely search for resonances in the generalized system.

Because near-Keplerian motion is under investigation, a natural analytical
convention to use is the osculating ellipse. The otherwise constant elements
of the ellipse for unperturbed motion become time-dependent osculating
elements with the perturbations. Each instantaneous position in the motion is
a point of an ellipse that is characterized by the osculating elements at 
that time. If at that time the perturbation were supressed, the motion
would follow the path of the osculating ellipse uniquely described by the 
instantaneous position and velocity. 
One can, of course, transform between coordinate systems.
The scheme for numerical work, for example, was done in polar coordinates. 
When one takes equation (28), applies the transformation to polar
coordinates
\begin{equation}
x=r\sin{\theta}, \ y=r\cos{\theta},
\end{equation}
and converts the second-order ODEs into equivalent first-order ODEs one
obtains the following system of equations:
\begin{equation}
\dot{r}=P_{r},
\end{equation}
\begin{equation}
\dot{\theta}=\frac{P_{\theta}}{r^{2}},
\end{equation}
\begin{equation}
\dot{P}_{r} = \frac{P_{\theta}^{2}}{r^{3}} - \frac{1}{r^{2}} +
\frac{1}{2}
\epsilon r [1+3\cos{(\Omega' t -2\theta)}] + \delta \frac{P_{r}}{r^{3}}
(\frac{32}{r} + 24 P_{r}^{2} + 144 \frac{P_{\theta}^{2}}{r^{2}}),
\end{equation}
\begin{equation}
\dot{P}_{\theta} = \frac{3}{2}\epsilon r^{2}\sin{( \Omega' t - 2\theta)}
+\delta \frac{P_{\theta}}{r^{3}} (\frac{24}{r} +108 P_{r}^{2} -
72\frac{P_{\theta}^{2}}{r^{2}}).
\end{equation}
Here $\Omega' = 2\Omega$; furthermore,
this polar scheme with the usual variables $r$ and $\theta$ gives rise to
the variables $P_{r}$ and $P_{\theta}$ defined above in (31) and (32) in this
first order form. $P_{\theta}$ represents the angular momentum of the two-body
system per reduced mass.

One can generate a series of numerical integrations in search of resonances
of various orders. Once the parameters that represent the amplitudes of
the tidal perturbation due to the presence of the third mass and gravitational 
radiation damping are chosen, one
can numerically generate an orbit starting from initial conditions. 
In anticipation of future analytical work to be done in Delaunay variables
(Wardell, to be published), results of the numerical integration are graphed
in terms of Delaunay variables. In figure 1, the orbit of the Delaunay action
$L$ is plotted versus time. The variable $L$ is directly related
to the semimajor axis by the relation $L=a^{1/2}$, where $a$ is the semimajor
axis of the relative orbit. The left side of the orbit makes a sharp descent
which accords with the expectation of semimajor axis decay due to the influence
of gravitational radiation reaction. However, the orbit commences to undergo
oscillations about an average value. A resonance occurs here. The amplitude
of the oscillations increase until the orbit falls out of resonance. This is
a nonlinear feature of the orbit. The orbit eventually falls out of the resonance
and gives way to a descent whose slope is more gradual than that of the initial
sharp descent due to the fact that the orbital eccentricity changes during
resonance as a consequence of the variation of orbital angular momentum.
The phenomena associated with this resonance are discussed in the next
section.

\section{RESONANCE}

Numerical experiments show that a $(1:1)$ resonance exists for the
system with the appropriate initial conditions. Checks on the
numerical integration have been performed to rule out the possibilty that
the effects seen are numerical artifacts.

The gravitational environment of the binary comes from two contributors:
the tidal influence of the third mass and the emission of 
gravitational waves.
These combined effects enter as the perturbations in the dynamical system.
The interplay of tidal energy input from the orbit around the third mass and 
the energy loss caused by the emission of gravitational waves comes to
a place of balance during a resonance. A resonance is characterized by
the resonance condition $m\omega = 2n \Omega$, where $\omega$ is the
angular frequency of the perturbed Kepler orbit and $\Omega$ is the
angular frequency of the binary's orbit around the third mass.
Because the radiation damping is a dissipative effect, decay of the
semimajor axis might be expected as the dominant perturbative effect. 
In fact, the example given by the Hulse-Taylor binary pulsar where this decay 
is predominantly observed
shows this effect quite well. However, in the presence of a third mass
resonance, as seen in this system, fixes 
the \textit{average} net flux of energy and gives rise to an interesting
nonlinear dynamical system whose 
perturbations of tidal interaction and gravitational radiation damping
offset each other on the average.
Changes in the orbital angular momentum, though, can accompany this fixed 
average energy that follows from the resonance condition.
The binary's orbit around the third mass can result in an increase in
its internal orbital angular momentum due to the tidal torque of the
external mass.
On the other hand, gravitational waves emitted from the binary carry with them 
angular momentum that decreases the orbital angular momentum in the binary.

\subsection{CAPTURE INTO RESONANCE}

Approaching from the left in Figure 1, the semimajor axis of the orbit steadily 
shrinks 
until it reaches the resonance manifold. The resonance manifold is defined 
by  $L=L_{*}$, where $L_{*}=\omega_{*}^{-1/3}$ and the resonance 
condition for a $(m:n)$ resonance fixes $\omega_{*}$ such that 
$\omega_{*}=2(n/m) \Omega$. This in turn fixes $L_{*}$.
Resonance capture is noteworthy because the orbit of the 
system passes through many resonances unimpeded. The integers $m$ and $n$ must 
be relatively 
prime, making $n/m$ a rational number; since the rational numbers are dense
with respect to the real numbers,
the orbit is always close to a resonance without being captured, except under
special circumstances. Numerical experiments show that capture more readily
occurs for low-lying resonances such as $(1:1)$ as illustrated in Figures 1-3.
Moreover, higher order resonances give rise to more chaotic structures in
their corresponding figures for $L$, $P_{\theta}$, and $e$ (Chicone, Mashhoon, and 
Retzloff, 1997b). Here $P_{\theta}$ is the specific orbital angular momentum
of the osculating ellipse and $e$ is its eccentricity defined by
$(1-P_{\theta}^{2}/L^{2})^{1/2}$.

\subsection{PASSAGE THROUGH RESONANCE}

When an orbit is captured into a resonance it enters the resonance manifold.
That is, the average $L$ value is fixed at $L_{*}$. The orbit then oscillates
about this $L_{*}$ value with increasing amplitude. Eventually, it falls out
of the resonance. While in resonance, the other action
variable $P_{\theta}$ can on average change significantly. 
This is shown in Figure 2, which
indicates that through the resonance $L$ oscillates about an average
value $L_{*}$, $P_{\theta}$ increases on average during this interval. 
Thus in the course of
resonance, the osculating ellipse associated with the orbit undergoes an
average change in eccentricity, as demonstrated in Figure 3.

\subsection{EXIT FROM RESONANCE}

The energy loss due to the emission of gravitational waves depends on the
eccentricity as $(1-e^{2})^{-7/2}$ (Landau and Lifshitz, 1971). 
Therefore, the decrease in this  orbital eccentricity
accounts for the apparent lower rate of energy loss once the orbit leaves the
resonance as in Figure 1. The anti-damping seen through the resonance 
eventually disrupts the resonance condition and leads to the orbit falling out 
of resonance. Where the orbit falls out of resonance the eccentricity is 
$e \simeq 0.73$ as compared to $e \simeq 0.85$ when the orbit was first 
captured into resonance.

An approximate analytic description of the phenomenon associated with resonance
is the subject of a future publication (Wardell, to be published).

\begin{figure}

\epsfig{file=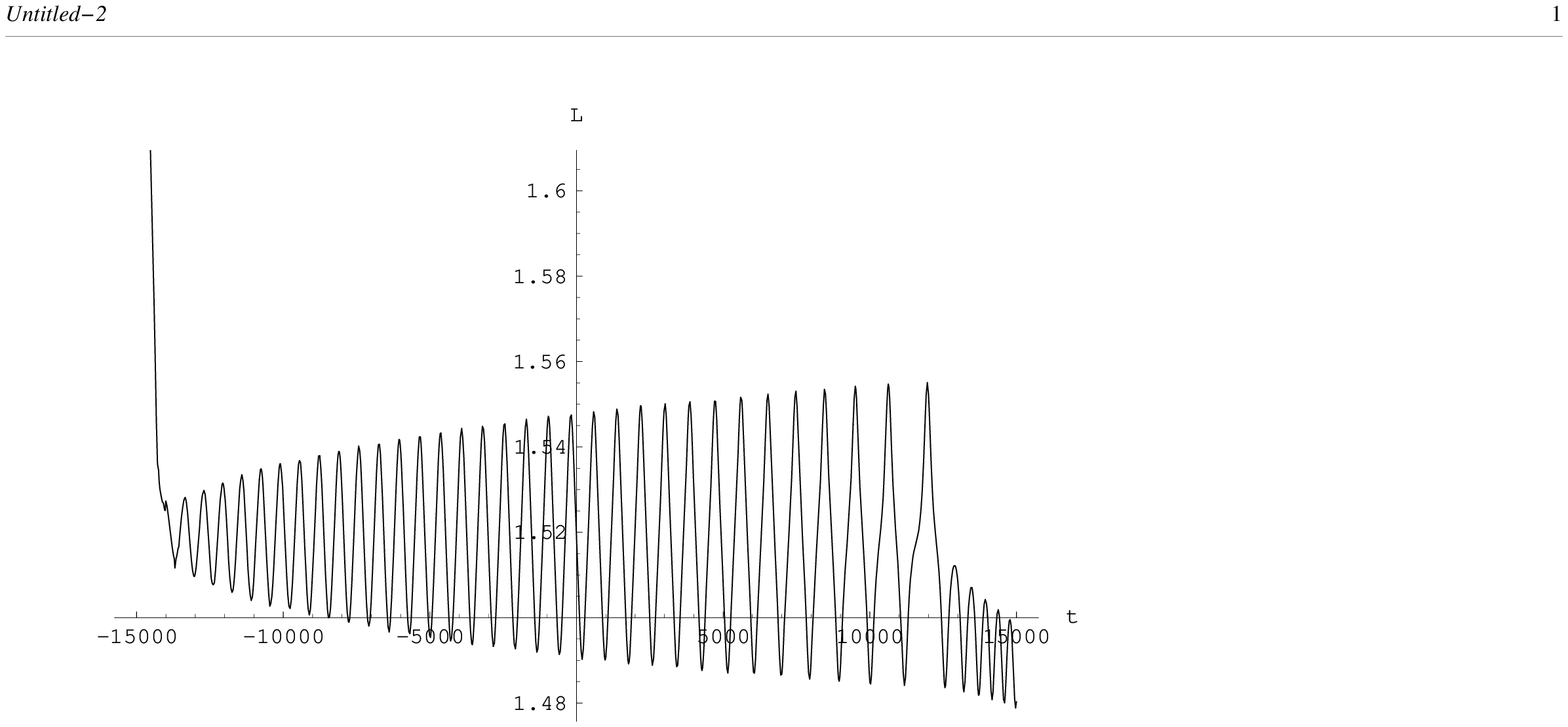,width=13cm,height=13cm,angle=-90,clip=,bbllx=50,bblly=95,
bburx=400,bbury=520}

\caption{This is a (1:1) resonance. The parameters are $\epsilon=0.00003$, 
$\delta/\epsilon=10^{-3}$, and $\Omega=0.14286$. The initial conditions at $t=0$
are $(r,\theta,P_{r},P_{\theta})=(1,0.57,0.75246,1).$}

\end{figure}
\begin{figure}

\epsfig{file=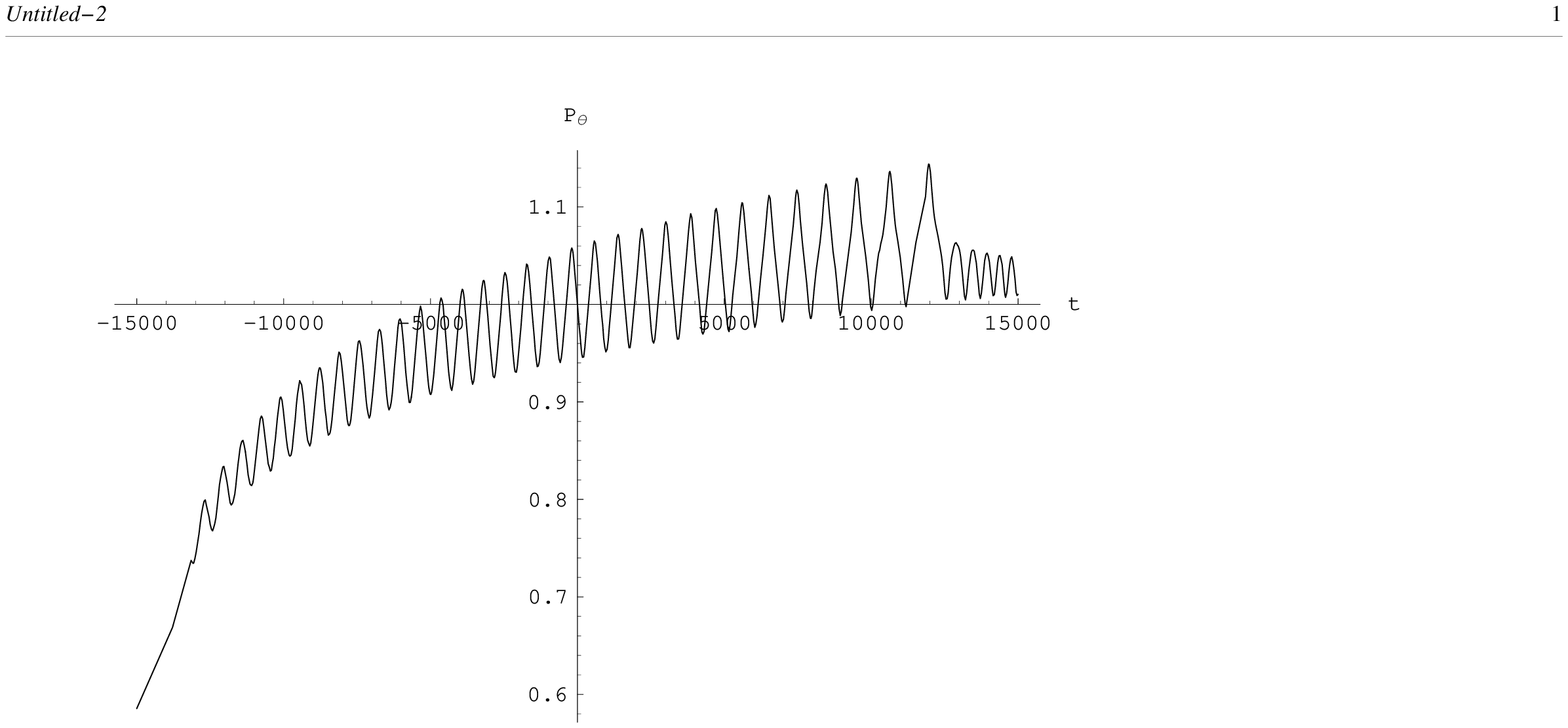,width=13cm,height=13cm,angle=-90,clip=,bbllx=50,bblly=95,
bburx=400,bbury=520}

\caption{Plot of $P_{\theta}$ versus time for the (1:1) resonance.}

\end{figure}

\begin{figure}

\epsfig{file=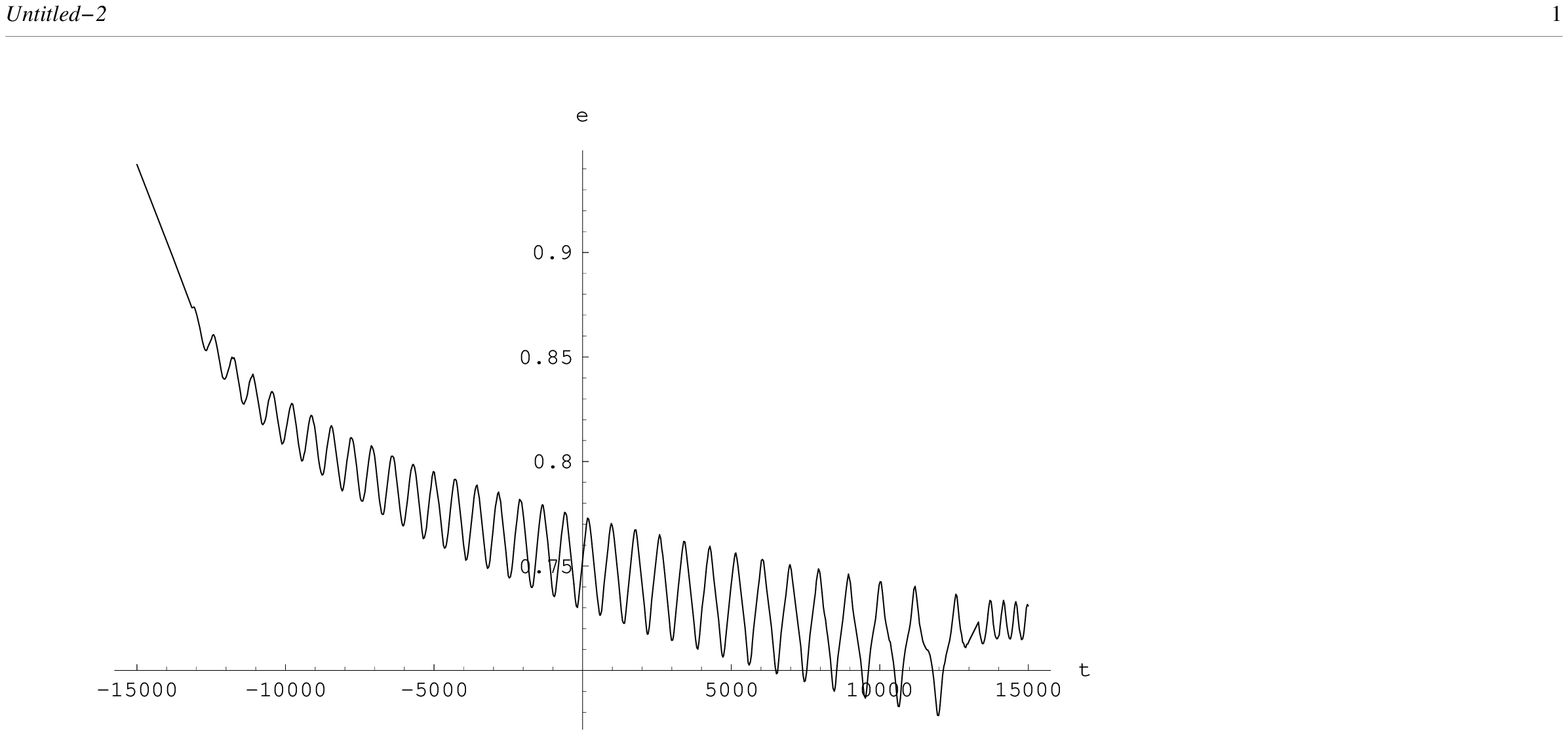,width=13cm,height=13cm,angle=-90,clip=,bbllx=50,bblly=95,
bburx=400,bbury=520}

\caption{Plot of eccentricity versus time for the (1:1) resonance.}

\end{figure}

\section{CONCLUSION}

The main thrust of the analysis in this paper has been to introduce and develop 
a model of a binary system that is under the influence of
gravitational radiation reaction and perturbed by a third body. The classical
case of three gravitating bodies is considered here when the damping force due
to the emission of gravitational waves by the system is included in the
Newtonian equations of motion. After the pertinent equations of motion
were derived, numerical analysis was done to explore important dynamics in
the system -- namely, resonance between the orbits of the relative and third
body motions. This would optimistically be an astronomically observable effect.

An analytical approach that would resolve details about this nonlinear system
would add to the confidence of the numerical results. What remains, to
augment the results of this work, is to pursue an analytical solution.
The previous work of Chicone, Mashhoon, and Retzloff (1996,1997) outlines a novel
averaging approach that looks into orbits near resonance. It was applied
to the case of a binary system perturbed by the emission and absorption of
gravitational waves. The model presented in this paper along with numerical
indication of the existence of resonances leads naturally to this analytical
approach for the model developed here (Wardell, to be published).

\vspace{.5cm}

\begin{flushleft}
\textbf{\Large{ACKNOWLEDGEMENTS}}\\
\vspace{.4cm}
I would like to thank B. Mashhoon for his indispensable help and guidance in 
this project. I would also like to thank B. DeFacio for generously offering his 
computer facilities.
\end{flushleft}

\begin{flushleft}
\textbf{\Large{REFERENCES}}\\
\vspace{.4cm}
Bennet D.P., et al., 1999, Nature, 402, 57\\
Blanchet, 2002, gr-gc/0202016, to appear in Living Reviews in Relativity\\
Cameron A.C.,2001, Physics World, 14, 1\\
Chicone C.,Mashhoon B.,Retzloff D.G., 1996a, Ann. Inst. 
Henri Poincar\'{e},\\ \hspace{.5cm} Phys. Th\'{e}or., 64, 87\\
Chicone C.,Mashhoon B.,Retzloff D.G., 1996b, J. Math. Phys., 37, 3997\\
Chicone C.,Mashhoon B.,Retzloff D.G., 1997a, Class. Quantum Grav.,
14, 699\\
Chicone C.,Mashhoon B.,Retzloff D.G., 1997b, Class. Quantum Grav., 
\\ \hspace{.5cm}14, 1831\\
Chicone C.,Kopeikin S.,Mashhoon B.,Retzloff D.G., 2001, Phys. Lett. A,
\\ \hspace{.5cm} 285,17\\
Landau L.D., Lifshitz E.M. 1971, The Classical Theory of Fields (Oxford:
\\ \hspace{.5cm}Pergamon Press)\\
Lyne Andrew G., Graham-Smith Francis 1998, Pulsar Astronomy (Cambridge:
\\ \hspace{.5cm} Cambridge University Press)\\
Schott G.A. 1912, Electromagnetic Radiation (London and New York:
\\ \hspace{.5cm}Cambridge University Press)\\
Stairs I.H., et al.,1989,Astrophys. J.,505,352
\end{flushleft}

\appendix

\renewcommand{\thesection}{APPENDIX \Alph{section}:}
\setcounter{equation}{0}
\renewcommand{\theequation}{\Alph{section}\arabic{equation}}
\section{THE SECOND SCENARIO}

Imagine a relativistic binary system consisting of masses $m_{1}$ and
$m_{2}$ and a distant third mass $m_{3}$ (with $m_{3}<<m_{1}$, and
$m_{3}<<m_{2}$) in a nearly circular orbit about the binary system
such that $r<<|\mathbf{x_{3}}|$. Starting from (1)-(4),
the equation of relative motion is

\begin{equation}
\frac{d^{2}r^{i}}{dt^{2}} + \frac{G Mr^{i}}{r^{3}}=
Gm_{3}\left(\frac{x_{3}^{i}-x_{2}^{i}}{|\mathbf{x}_{3}-\mathbf{x}_{2}|^{3}}
-\frac{x_{3}^{i}-x_{1}^{i}}
{|\mathbf{x}_{3}-\mathbf{x}_{1}|^{3}}\right)- \frac{2 G}{15 c^{5}} 
\frac{d^{5} D_{ij}}{dt^{5}}r^{j}.
\end{equation}

Writing $x^{i}_{1}=X^{i}-\gamma_{2}r^{i}$ and
$x^{i}_{2}=X^{i}+\gamma_{1}r^{i}$ as before, and
recalling that $X^{i}= -m_{3}x^{i}_{3}/M$, we find

\begin{equation}
\frac{x_{3}^{i}-x_{2}^{i}}{|\mathbf{x}_{3}-\mathbf{x}_{2}|^{3}}=
\frac{(1+\frac{m_{3}}{M})x_{3}^{i}-\gamma_{1}r^{i}}
{|(1+\frac{m_{3}}{M})\mathbf{x}_{3}-\gamma_{1}\mathbf{r}|^{3}},
\end{equation}

\begin{equation}
\frac{x_{3}^{i}-x_{1}^{i}}{|\mathbf{x}_{3}-\mathbf{x}_{1}|^{3}}=
\frac{(1+\frac{m_{3}}{M})x_{3}^{i}+\gamma_{2}r^{i}}
{|(1+\frac{m_{3}}{M})\mathbf{x}_{3}+\gamma_{2}\mathbf{r}|^{3}}.
\end{equation}
Using $m_{3}<<M$ and neglecting $m_{3}/M$ in comparison to unity
meanwhile expanding (A2) and (A3) as in equations (18) and (19) based
on the fact that $r<<|\mathbf{x_{3}}|$, one finds that to lowest
order the result is

\begin{equation}
\frac{d^{2}r^{i}}{dt^{2}} = - \frac{GM}{r^{3}}r^{i} 
-\frac{G m_{3}}{|\mathbf{x_{3}}|^{3}} K_{ij} r^{j}
- \frac{2 G}{15 c^{5}} \frac{d^{5} D_{ij}}{dt^{5}} r^{j},
\end{equation}
where $D_{ij} \simeq \mu(3r^{i}r^{j}-\delta_{ij}r^{2})$ once the condition

\begin{equation}
\frac{m_{3}}{\mu} \left( \frac{r}{|\mathbf{x_{3}}|} \right)^{11/2}
<< 1
\end{equation}
is imposed. In equation (A4), $K_{ij}$ is given by

\begin{equation}
K_{ij}=\delta_{ij} - 3 \frac{x_{3}^{i}x_{3}^{j}}{|\mathbf{x_{3}}|^{2}}.
\end{equation}
It is clear that this scenario, suitably interpreted, gives essentially
the same equation of relative motion that is given in equation (27).

\setcounter{equation}{0}
\section{ITERATIVE REDUCTION}

To obtain the desired approximation of the radiation reaction force,
one makes successive differentiations of the quadrupole moment tensor
and substitutes the Keplerian equation of motion whenever the second
time derivative of position appears. The reduced radiation reaction is:
\begin{equation}
R^{i}=\frac{d^{5}\tilde{D}_{ij}}{dt^{5}}r^{j}.
\end{equation}
The scaled Kepler equation of motion is:
\begin{equation} 
\frac{d^{2}r^{i}}{dt^{2}}=-\frac{r^{i}}{r^{3}}. 
\end{equation}
For brevity, $\dot{\mathbf{r}}$ will be represented by $\mathbf{v}$.
One starts with the quadrupole moment tensor
\begin{equation} 
\tilde{D}_{ij}=3r^{i}r^{j}-\delta_{ij}r^{2}
\end{equation}
and proceeds to take time derivatives:
\begin{equation} 
\frac{d \tilde{D}_{ij}}{dt} = 3\left(r^{i}v^{j} + v^{i}r^{j}\right) -
\delta_{ij} \left(2
\textbf{r}\cdot\textbf{v}\right),
\end{equation}

\begin{equation}
\frac{d^{2}\tilde{D}_{ij}}{dt^{2}} = 3\left(r^{i}\dot{v}^{j} +
\dot{v}^{i}r^{j} + 2
v^{i}
v^{j}\right) - 2\delta_{ij}\left( \textbf{v}^{2} + \textbf{r} \cdot 
\dot{\textbf{v}}\right).
\end{equation}
Where $\dot{v}^{i}$ occurs, one
substitutes the
unperturbed Kepler equation of motion to arrive
at the following
\begin{equation}
\frac{d^{2}\tilde{D}_{ij}}{dt^{2}} = 6\left(
-\frac{r^{i}r^{j}}{r^{3}}
 + v^{i}v^{j}\right)-2\delta_{ij}\left(v^{2}-\frac{1}{r}\right).
\end{equation}
Subsequent differentiations and substitutions yield the following expressions:
\begin{equation}
\frac{d^{3}
\tilde{D}_{ij}}{dt^{3}}=-\frac{12}{r^{3}}\left(v^{i}r^{j}+r^{i}v^{j}\right)
+18 \frac{r^{i}r^{j}}{r^{4}} \dot{r} -
\delta_{ij}\left(-2\frac{\dot{r}}{r^{2}}\right), 
\end{equation}

\begin{equation}
\frac{d^{4}\tilde{D}_{ij}}{dt^{4}}=
-\frac{24v^{i}v^{j}}{r^{3}}+\frac{54(v^{i}r^{j}+r^{i}v^{j})}
{r^{4}}\dot{r}+\left(-90\dot{r}
+18v^{2}+\frac{6}{r}\right)\frac{r^{i}r^{j}}{r^{5}}
\end{equation}
\[-\delta_{ij}\left(\frac{6}{r^{3}}\dot{r}^{2} -
\frac{2v^{2}}
{r^{3}} + \frac{2}{r^{4}}\right),\]  

\begin{equation}
\frac{d^{5} \tilde{D}_{ij}}{dt^{5}}=
\frac{180v^{i}v^{j}}{r^{4}}\dot{r}+
\left(-\frac{24}{r}-360 \dot{r}^{2} + 72v^{2}\right)
\left(\frac{v^{i}r^{j}+r^{i}v^{j}}{r^{5}}\right)                              
\end{equation}
\[+\left(630 \dot{r}^{2}-270
v^{2}\right)\frac{r^{i}r^{j}}{r^{6}} \dot{r}
-\delta_{ij}\left(-\frac{30 \dot{r}^{3}}{r^{4}}+
\frac{18v^{2}\dot{r}}{r^{4}}
-\frac{16 \dot{r}}{r^{5}}\right).\] 
Finally, when one contracts expression (B9) with $r^{j}$ to compute the desired
reduced radiation reaction expression as prescribed by (B1), 
one finds equation (29).

\end{document}